\documentclass[twocolumn,longbibliography]{revtex4}
\usepackage[dvips]{graphicx}
\usepackage{float, color,array, multirow}

\begin{document}

\title{Hedgehog-lattice spin texture in classical Heisenberg antiferromagnets on the breathing pyrochlore lattice}

\author{Kazushi Aoyama$^1$ and Hikaru Kawamura$^2$}

\date{\today}

\affiliation{${}^{1}$Department of Earth and Space Science, Graduate School of Science, Osaka University, Osaka 560-0043, Japan \\
${}^{2}$Toyota Physical and Chemical Research Institute, Aichi, 480-1118, Japan
}

\begin{abstract}
The hedgehog lattice, a three-dimensional periodic array of magnetic monopoles and antimonopoles, is known to be realized in the presence of the Dzyaloshinskii-Moriya (DM) interaction. Here, we demonstrate by means of Monte Carlo simulations that the hedgehog lattice is induced by not the DM interaction but frustration in classical Heisenberg antiferromagnets on the breathing pyrochlore lattice. In the model, the breathing bond-alternation is characterized by the ratio of the nearest-neighbor (NN) antiferromagnetic exchange interaction for large tetrahedra to that for small ones, $J_1'/J_1$. A quadruple-${\bf q}$ state with the ordering vector of ${\bf q}=(\pm\frac{1}{2},\pm\frac{1}{2},\pm\frac{1}{2})$, which is realized for a large third NN antiferromagnetic interaction along the bond direction $J_3$, turns out to become the hedgehog-lattice state in the breathing case of $J_1'/J_1 <1$, while in the uniform case of $J_1'/J_1 =1$, it is a collinear state favored by thermal fluctuations. It is also found that in a magnetic field, the structure of the $(\frac{1}{2},\frac{1}{2},\frac{1}{2})$ hedgehog lattice is changed from cubic to tetragonal, resulting in a nonzero net spin chirality which in a metalilc system, should yield a characteristic topological Hall effect.
\end{abstract}

\maketitle
\section{introduction}
Topological spin textures such as the magnetic skyrmion and its three-dimensional analogue, the magnetic hedgehog, have attracted much attention because of their possible applications to spin-electronic devices \cite{SkX_review_Nagaosa-Tokura_13, SkX_Koshibae_15, MnGe_Fujishiro_18}. Of recent particular interest are periodic arrays of these topological objects. It is nowadays widely accepted that the Dzyaloshinskii-Moriya (DM) interaction is a key ingredient for crystalline orders of the skyrmions \cite{SkX_review_Nagaosa-Tokura_13, MnSi_Kadowaki_82, MnSi_Muhlbauer_09, MnSi_Neubauer_09, FeCoSi_Yu_10, FeGe_Yu_11, Fefilm_Heinze_11, Cu2OSeO3_Seki_12, CoZnMn_Tokunaga_15, GaV4S8_Kezsmarki_15, GaV4Se8_Fujima_17, GaV4Se8_Bordacs_17, VOSe2O5_Kurumaji_17,AntiSkX_Nayak_17, EuPtSi_Kakihana_18, EuPtSi_Kaneko_19, SkX_Robler_06, SkX_Yi_09, SkX_Buhrandt_13} and the hedgehogs \cite{MnGe_Kanazawa_11, MnGe_Kanazawa_12, MnGe_Shiomi_13, MnGe_Tanigaki_15, MnGe_Kanazawa_16, MnGe_Kanazawa_17, MnSiGe_Fujishiro_19, MnGe_Fujishiro_20, Hedgehog_MFtheory_Binz_prl06, Hedgehog_MFtheory_Binz_prb06, Hedgehog_MFtheory_Binz_08, Hedgehog_MFtheory_Park_11, Hedgehog_MCtheory_Yang_16, Hedgehog_MCtheory_Okumura_19}. Besides, it has been realized since the pioneering work by Okubo, Chung, and Kawamura \cite{SkX_Okubo_12} that even in centrosymmetric magnets without the DM interaction, the skyrmion lattice appears as a result of competitions between exchange interactions, i.e., magnetic frustration \cite{SkX_Leonov_15,SkX_Lin_16,SkX_Hayami_16,SkXimp_Hayami_16,SkX_Lin_18,SkX_Shimokawa_19,RKKY,Gd2PdSi3_Kurumaji_19,GdRu2Si2_Khanh_20}. Concerning the hedgehog lattice, however, the counter ordering mechanism other than the DM interaction has not been reported so far. In this paper, we will theoretically demonstrate that the hedgehog lattice can be induced by the frustration in classical Heisenberg antiferromagnets on the breathing pyrochlore lattice.

The magnetic skyrmion and hedgehog are spin textures both characterized by the integer topological charge which corresponds to, in units of $4\pi$, the total solid angle subtended by all the spins involved. The distinction between the two is in the real-space structure. In contrast to the skyrmion in two dimensions, the hedgehog extends over the three-dimensional real space, so that it has a sigular point in the texture center. In this respect, the hedgehog is sometimes called the magnetic monopole. Noting that the solid angle $\Omega_{ijk}$ for three spins ${\bf S}_i$, ${\bf S}_j$, and ${\bf S}_k$ is related with the scalar spin chirality $\chi_{ijk}={\bf S}_i \cdot ({\bf S}_j \times {\bf S}_k)$ via $\Omega_{ijk}=2\tan^{-1}\big[\frac{\chi_{ijk}}{1+{\bf S}_i\cdot{\bf S}_j +{\bf S}_i\cdot{\bf S}_k+{\bf S}_j\cdot{\bf S}_k} \big]$ \cite{SolidAngle_OOsterom_83}, the hedgehog can be understood as a topologically protected chirality object. 

The hedgehog lattice is an alternating periodic array of the hedgehogs (monopoles) and anti-hedgehogs (anti-monopoles) each having positive or negative nonzero scalar spin chirality. In other words, it is a long-range order (LRO) of the chirality. When viewed as a LRO of spin itself, the hedgehog lattice is a multiple-${\bf q}$ state described by more than one ordering-wave-vectors ${\bf q}$'s \cite{MnGe_Kanazawa_11,MnGe_Kanazawa_12, MnGe_Tanigaki_15, MnGe_Kanazawa_16, MnGe_Kanazawa_17, MnSiGe_Fujishiro_19, Hedgehog_MFtheory_Binz_prl06, Hedgehog_MFtheory_Binz_prb06, Hedgehog_MFtheory_Binz_08, Hedgehog_MFtheory_Park_11, Hedgehog_MCtheory_Yang_16, Hedgehog_MCtheory_Okumura_19}. A prominent aspect of the hedgehog lattice is the Hall effect of spin chirality origin \cite{THE_Tatara_02}, i.e., the so-called topological Hall effect. In the non-centrosymmetric magnets having the DM interaction MnSi$_{1-x}$Ge$_x$ with $0.25<x$, the occurrence of the hedgehog lattice has been evidenced by the topological Hall effect together with the observation of multiple-${\bf q}$ Bragg reflections \cite{MnGe_Kanazawa_11, MnGe_Kanazawa_12, MnGe_Shiomi_13, MnGe_Tanigaki_15, MnGe_Kanazawa_16, MnGe_Kanazawa_17, MnSiGe_Fujishiro_19}. Recently, the hedgehog lattice phase was found also in the DM-free centrosymmetric magnet SrFeO$_3$ \cite{SFO_Ishiwata_11, SFO_Ishiwata_20}, but the ordering mechanism is still unclear and no other candidate compounds have been reported so far. In this work, towards the exploration of new classes of magnets hosting the hedgehog-lattice spin texture, we theoretically investigate effects of the frustration which in two dimensions, induces the skyrmion lattice \cite{SkX_Okubo_12}. A typical example of frustrated three-dimensional systems would be classical Heisenberg antiferromagnets on the pyrochlore lattice.

\begin{figure}[t]
\begin{center}
\includegraphics[scale=0.75]{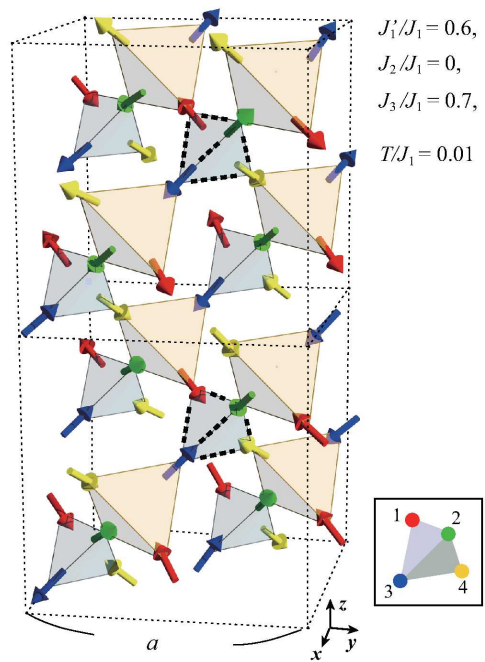}
\caption{Hedgehog-lattice spin texture at zero field on the breathing pyrochlore lattice, where the spin configuration is obtained at $T/J_1=0.01$ in the Monte Carlo (MC) simulation for the Hamiltonian (\ref{eq:Hamiltonian}) with $J'_1/J_1=0.6$, $J_2/J_1=0$ and $J_3/J_1=0.7$. Red, green, blue, and yellow arrows represent spins on the corners 1, 2, 3, and 4 of the small tetrahedra shown in the inset, respectively. The small tetrahedron at the center of the cubic unit cell is outlined by black dots. The upper and lower central small tetrahedra correspond to the hedgehog (monopole) and the anti-hedgehog (anti-monopole), respectively (see the main text in Sec. III). \label{fig:hedgehog_snapshot}}
\end{center}
\end{figure}

The pyrochlore lattice is a three-dimensional network of corner-sharing tetrahedra. In the presence of not only the nearest-neighbor(NN) antiferromagnetic exchange interaction $J_1$ but also the third NN one along the bond directions $J_3$, it turns out that classical Heisenberg spins are ordered into a quadruple-${\bf q}$ state with the four ordering vectors ${\bf q}=(\frac{1}{2}, \frac{1}{2}, \frac{1}{2}), \, (-\frac{1}{2}, \frac{1}{2}, \frac{1}{2}), \, (\frac{1}{2}, -\frac{1}{2}, \frac{1}{2})$, and $(\frac{1}{2}, \frac{1}{2}, -\frac{1}{2})$ in units of $\frac{2\pi}{a}$ in the cubic basis \cite{Site_AK_16,Site_AK_19,J1J2J3_Mizoguchi_18} and that the spins are collinearly aligned due to the effect of thermal fluctuations. Since the above $J_1$-$J_3$ pyrochlore antiferromagnet can potentially host the hedgehog lattice in the sense that the ordered phase is the multiple-${\bf q}$ state, we try to suppress the spin collinearity. For this purpose, we consider the so-called breathing pyrochlore lattice consisting of an alternating array of small and large tetrahedra (see Fig. \ref{fig:hedgehog_snapshot}) \cite{BrPyro_Okamoto_13, Site_AK_19}.

In the breathing pyrochlore lattice, the bond alternation should be reflected in different values of the NN interactions on small and large tetrahedra, $J_1$ and $J_1'$, so that its strength could be measured by the ratio $J_1'/J_1$. In the different context of the spin-lattice coupling, the effect of the breathing lattice structure has already been examined: within the collinear-spin manifold, the fully ordered quadruple-${\bf q}$ state with ${\bf q}=(\pm\frac{1}{2}, \pm\frac{1}{2}, \pm\frac{1}{2})$ is changed into a partially disordered one for smaller values of $J_1'/J_1$ \cite{Site_AK_19}. This indicates that for the present Heisenberg spins, the thermally-driven collinear $(\frac{1}{2}, \frac{1}{2}, \frac{1}{2})$ state would get unstable with decreasing $J_1'/J_1$, possibly leading to the hedgehog lattice. 
As demonstrated in Fig. \ref{fig:hedgehog_snapshot}, this is actually the case. One can see from Fig. \ref{fig:hedgehog_snapshot} that the quadruple-${\bf q}$ $(\frac{1}{2}, \frac{1}{2}, \frac{1}{2})$ state on the breathing pyrochlore lattice is nothing but the hedgehog lattice (see Secs. II and III for details). This $(\frac{1}{2}, \frac{1}{2}, \frac{1}{2})$ hedgehog lattice appears as a result of the frustration in the absence of the DM interaction. In this paper, we
further clarify that the hedgehog lattice in a magnetic field possesses a nonzero net spin chirality which should yield emergent phenomena such as the topological Hall effect.

This paper is organized as follows: In Sec. II, we introduce the spin Hamiltonian and discuss its ordering properties based on analytical calculations. The results of our MC simulations will be shown in Sec. III. In Sec. IV, we will discuss the effect of a magnetic field and demonstrate the emergence of the net spin chirality which results in an unconventional topological Hall effect quite different from the DM-induced one. The origin of this field-induced topological Hall effect will be addressed in Sec. V. We end the paper with summary and discussions in Sec. VI.

\section{Model Hamiltonian and its basic ordering properties}
\subsection{Model}
The spin Hamiltonian we consider is given by
\begin{eqnarray}\label{eq:Hamiltonian}
{\cal H} &=& J_1 \sum_{\langle i,j \rangle_S} {\bf S}_i \cdot {\bf S}_j + J_1' \sum_{\langle i,j \rangle_L} {\bf S}_i \cdot {\bf S}_j \nonumber\\
&& +  J_2 \sum_{\langle \langle i,j \rangle \rangle} {\bf S}_i \cdot {\bf S}_j + J_{3} \sum_{\langle \langle \langle i,j \rangle \rangle \rangle} {\bf S}_i \cdot {\bf S}_j,
\end{eqnarray} 
where ${\bf S}_i$ is the classical Heisenberg spin at the site $i$, $\langle \rangle_{S(L)}$, $\langle \langle i,j \rangle \rangle$, and $\langle \langle \langle i,j \rangle \rangle \rangle$ denote the summations over site pairs on the small (large) tetrahedra, the second NN pairs, and the third NN pairs along the bond direction, respectively. As explained in Sec. I, the breathing bond-alternation is characterized by the ratio of the NN interaction for large tetrahedra to that for small ones, $J_1'/J_1$, and the third NN antiferromagnetic interaction along the bond direction $J_3$ is crucial for the occurrence of the hedgehog lattice. Although the second NN interaction $J_2$ is not essential, we have included $J_2$ for completeness.
When we discuss the in-field properties of the system in Sec. IV, we include the additional Zeeman term $-H\sum_i S_{i,z}$ in the Hamiltonian (\ref{eq:Hamiltonian}), where $H$ is the intensity of a magnetic field.
Throughout this paper, we take the cubic unit cell of edge length $a=1$ shown in Fig. \ref{fig:hedgehog_snapshot}, and thus, the total number of spin $N$ and the linear system size $L$ is related via $N=16L^3$. 

\subsection{Effect of the breathing lattice structure on the spin ordering}
\begin{figure}[t]
\begin{center}
\includegraphics[scale=0.6]{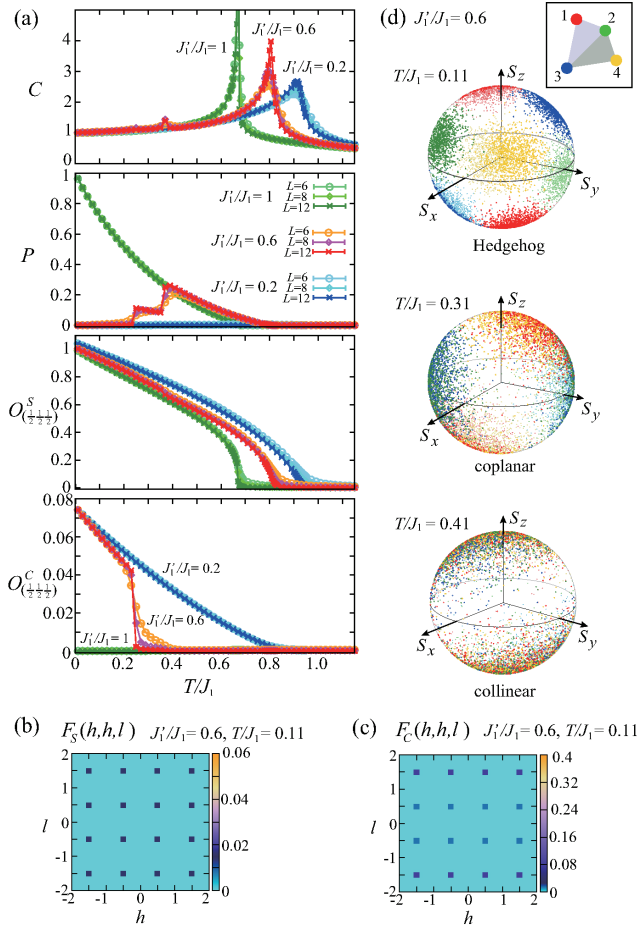}
\caption{MC results obtained for $J_2/J_1=0$ and $J_3/J_1=0.7$. Note that a magnetic field $H$ is not applied. (a) The temperature dependence of the specific heat $C$ (the first panel from the top), the spin collinearity $P$ (the second one), and the average spin and chirality Bragg intensities $O^S_{(\frac{1}{2}\frac{1}{2}\frac{1}{2})}$ and $O^C_{(\frac{1}{2}\frac{1}{2}\frac{1}{2})}$ (the third and fourth ones). Greenish, reddish, and bluish colored symbols denote the data for $J'_1/J_1=1$, $0.6$, and $0.2$, respectively. (b)-(e) The results obtained in the ordered phases for $J'_1/J_1=0.6$. (b) and (c) Spin and chirality structure factors $F_S({\bf q})$ and $F_C({\bf q})$ in the $(h,h,l)$ plane at $T/J_1=0.11$. (d) Snapshots of spins mapped onto a unit sphere in the hedgehog-lattice phase at $T/J_1=0.11$ (top), the coplanar phase at $T/J_1=0.31$ (middle), and the collinear phase at $T/J_1=0.41$ (bottom), where red, green, blue, and yellow dots represent spins on the corners 1, 2, 3, and 4 of the small tetrahedra shown in the inset, respectively. \label{fig:hedgehog}}
\end{center}
\end{figure}

To get insight into roles of the breathing bond-alternation, we first discuss ordering properties of this model in the mean-field (MF) approximation \cite{Reimers_MF, Okubo_pyro}. Introducing the Fourier transform ${\bf S}_i=\sum_{\bf q}{\bf S}^\alpha_{\bf q}\exp(i{\bf q}\cdot {\bf r}_i)$ with the site index $i=(\alpha, {\bf r}_i)$, we can rewrite ${\cal H}$ into the following form
\begin{equation}\label{eq:H_Ising_FT}
{\cal H} = \frac{N}{8}\sum_{\bf q} \sum_{\alpha,\beta=1}^4 J^{\alpha\beta}({\bf q}) {\bf S}^\alpha_{\bf q} \cdot {\bf S}^\beta_{-{\bf q}} + const,
\end{equation}
where the sublattice indices $\alpha=1, \, 2, \, 3, \, 4$ correspond to the four corners of a tetrahedron [see the insets of Figs. \ref{fig:hedgehog_snapshot} and \ref{fig:hedgehog} (d)]. By analyzing the lowest eigen value of the 4-by-4 matrix $J^{\alpha\beta}({\bf q})$, $\lambda_{\bf q}$, we find that for various sets of parameters we searched, possible ordering wave-vectors are $\frac{2\pi}{a}(\pm \frac{1}{2}, \pm \frac{1}{2}, \pm \frac{1}{2}), $$\frac{2\pi}{a}(\pm 1, \pm 1,\pm 1)$, and incommensurate ones (see Fig. \ref{fig:Jq_Paradep} in Appendix A). For large $J_{3}$, $\lambda_{\bf q}$ is obtained at ${\bf Q}_1=\frac{2\pi}{a}(\frac{1}{2},\frac{1}{2},\frac{1}{2})$, ${\bf Q}_2=\frac{2\pi}{a}(-\frac{1}{2},\frac{1}{2},\frac{1}{2})$, ${\bf Q}_3=\frac{2\pi}{a}(\frac{1}{2},-\frac{1}{2},\frac{1}{2})$, and ${\bf Q}_4=\frac{2\pi}{a}(\frac{1}{2},\frac{1}{2},-\frac{1}{2})$ as $\lambda_{{\bf Q}_i} = J_1+ J_1'-2J_3 - \sqrt{3(J_1- J_1')^2 + (J_1+ J_1'+4J_{3})^2}$. The associated eigen vectors are given by
\begin{eqnarray}\label{eq:Uq}
{\bf U}_{{\bf Q}_1} &=& c_N(-i \varepsilon, \, 1, \, -i\varepsilon, \, -i\varepsilon), \nonumber\\
{\bf U}_{{\bf Q}_2} &=& c_N(i \varepsilon, \, i\varepsilon, \, 1, \, i\varepsilon ),  \nonumber\\
{\bf U}_{{\bf Q}_3} &=& c_N(i \varepsilon, \, i\varepsilon, \, i\varepsilon, \, 1), \nonumber\\
{\bf U}_{{\bf Q}_4} &=& c_N(1, \, i \varepsilon, \, i\varepsilon, \, i\varepsilon )
\end{eqnarray}
with the normalization factor $c_N=1/\sqrt{1+3\varepsilon^2}$ and the dimensionless coefficient
\begin{equation}
\varepsilon = \frac{J_1+J'_1+4J_3}{3(J_1-J'_1)}\bigg[ \sqrt{1 + 3 \Big( \frac{J_1-J'_1}{J_1+J'_1+4J_3}\Big)^2 } -1 \bigg].
\end{equation}
$\varepsilon$ is zero for the uniform pyrochlore lattice with $J'_1/J_1=1$, and gradually increases with increasing the breathing alternation (decreasing $J'_1/J_1$). Note that $J_2$ does not show up in $\lambda_{{\bf Q}_\mu}$ and ${\bf U}_{{\bf Q}_\mu}$, and thus, irrelevant for the ordering properties.

In the $(\frac{1}{2},\frac{1}{2},\frac{1}{2})$ ordered phase, the mean field $\langle {\bf S}^\alpha_{\bf q}\rangle$ is described as
\begin{equation}\label{eq:S_q} 
\langle {\bf S}^\alpha_{\bf q}\rangle= \sum_{\mu=1}^4\big( U^\alpha_{{\bf Q}_\mu}\mbox{\boldmath $\Phi$}_{{\bf Q}_\mu} \delta_{{\bf q},{\bf Q}_\mu} + [U^\alpha_{{\bf Q}_\mu}]^\ast \mbox{\boldmath $\Phi$}_{{\bf Q}_\mu}^\ast \delta_{{\bf q},-{\bf Q}_\mu} \big),
\end{equation}
where $U^\alpha_{{\bf Q}_\mu}$ represents the $\alpha$th component of ${\bf U}_{{\bf Q}_\mu}$. In the uniform case of $J'_1/J_1 =1$ and thereby $\varepsilon=0$, all the ${\bf U}_{{\bf Q}_\mu}$'s are orthogonal to one another, so that $\mbox{\boldmath $\Phi$}_{{\bf Q}_1}$, $\mbox{\boldmath $\Phi$}_{{\bf Q}_2}$, $\mbox{\boldmath $\Phi$}_{{\bf Q}_3}$, and $\mbox{\boldmath $\Phi$}_{{\bf Q}_4}$ correspond to the polarization vectors of the spins belonging to sublattice 2, 3, 4, and 1, respectively. 
In other words, to make spins exist at each sublattice, the ordered phase should be the quadruple-${\bf q}$ state involving all the $\mbox{\boldmath $\Phi$}_{{\bf Q}_\mu}$'s, and thus, other multiple-${\bf q}$ states such as single-${\bf q}$, double-${\bf q}$, and triple-${\bf q}$ states are not realized in the present system. Then, the Ginzburg-Landau (GL) free energy in the form expanded with respect to $\mbox{\boldmath $\Phi$}_{{\bf Q}_\mu}$ is obtained as 
${\cal F}_{\rm GL}/(N/4) = f_2 + f_4 + \delta f_4$ with  
\begin{eqnarray}\label{eq:GL}
&& f_2 = 2\big[3T + \lambda_{{\bf Q}_\mu} \big] \sum_{\mu=1}^4 |\mbox{\boldmath $\Phi$}_{{\bf Q}_\mu}|^2, \\
&& f_4 = \frac{9T}{20} \, A_1 \sum_{\mu=1}^4 \Big( |\mbox{\boldmath $\Phi$}_{{\bf Q}_\mu} \cdot \mbox{\boldmath $\Phi$}_{{\bf Q}_\mu}|^2 + 2 |\mbox{\boldmath $\Phi$}_{{\bf Q}_\mu}|^4 \Big), \nonumber\\
&& \delta f_4 = \frac{9T}{20}\bigg[ A_2 \sum_{\mu<\nu} \Big( |\mbox{\boldmath $\Phi$}_{{\bf Q}_\mu}|^2 |\mbox{\boldmath $\Phi$}_{{\bf Q}_\nu}|^2 + \sum_{\varepsilon_s=\pm 1} |\mbox{\boldmath $\Phi$}_{{\bf Q}_\mu} \cdot \mbox{\boldmath $\Phi$}_{\varepsilon_s{\bf Q}_\nu}|^2 \Big)  \nonumber\\
&& - i \, A_3 \Big( \Big\{ \sum_{\mu\neq \nu \neq \rho \neq 1}(\mbox{\boldmath $\Phi$}_{-{\bf Q}_1} \cdot \mbox{\boldmath $\Phi$}_{{\bf Q}_\mu})(\mbox{\boldmath $\Phi$}_{{\bf Q}_\nu} \cdot \mbox{\boldmath $\Phi$}_{{\bf Q}_\rho})\Big\} - \Big\{ c.c. \Big\} \Big) \bigg], \nonumber\\
&& A_1 = \frac{2(1+3\varepsilon^4)}{(1+3\varepsilon^2)^2}, \, A_2 = \frac{16\varepsilon^2(1+\varepsilon^2)}{(1+3\varepsilon^2)^2} \nonumber, A_3 = \frac{16\varepsilon^3}{(1+3\varepsilon^2)^2} \nonumber.
\end{eqnarray}
Only in the breathing case of $J'_1/J_1 <1$, $\varepsilon$ is nonzero and thus, $\delta f_4$ is active. The most stable spin configuration is obtained by minimizing ${\cal F}_{\rm GL}$ under the constraint ${\overline S}^2 = 2\sum_{\mu=1}^4 |\mbox{\boldmath $\Phi$}_{{\bf Q}_\mu}|^2$. 
Noting that in the uniform case, each $\mbox{\boldmath $\Phi$}_{{\bf Q}_\mu}$ determines the spin orientation axis ${\hat P}_\mu$ at each sublattice, we assume that even in the breathing case, $\mbox{\boldmath $\Phi$}_{{\bf Q}_\mu}$ takes the form of
$\mbox{\boldmath $\Phi$}_{{\bf Q}_\mu} = \frac{\overline S}{\sqrt{8}} e^{i \theta_\mu} {\hat P}_\mu$.
Then, we have
\begin{eqnarray}\label{eq:GL_final}
&& f_2 = \big[3T + \lambda_{{\bf Q}_\mu} \big] {\overline S}^2, \nonumber\\
&& f_4 = \frac{27T}{160} \, A_1 \, {\overline S}^4, \nonumber\\
&& \delta f_4 = \frac{9T}{640}{\overline S}^4 \bigg[ A_2 \Big\{ 3 + ({\hat P}_1\cdot{\hat P}_2)^2 + ({\hat P}_1\cdot{\hat P}_3)^2 \nonumber\\
&& + ({\hat P}_1\cdot{\hat P}_4)^2 + ({\hat P}_2\cdot{\hat P}_3)^2 + ({\hat P}_2\cdot{\hat P}_4)^2 + ({\hat P}_3\cdot{\hat P}_4)^2 \Big\} \nonumber\\
&&  - 2A_3 \sin(\theta_1-\theta_2-\theta_3-\theta_4) \Big\{ ({\hat P}_1\cdot{\hat P}_2)({\hat P}_3\cdot{\hat P}_4) \nonumber\\
&& + ({\hat P}_1\cdot{\hat P}_3)({\hat P}_2\cdot{\hat P}_4) + ({\hat P}_1\cdot{\hat P}_4)({\hat P}_2\cdot{\hat P}_3) \Big\}  \bigg]. 
\end{eqnarray}

In the uniform case of $J'_1/J_1=1$, $\delta f_4=0$, so that the relation among different $\theta_\mu$ and ${\hat P}_\mu$ cannot be determined. In other words, spins at different sublattices are uncorrelated at the MF level. As will be shown below, however, thermal fluctuations induce the collinear inter-sublattice correlation. In the breathing case of $J'_1/J_1 <1$, $\delta f_4$ is active and the minimization condition for $\delta f_4$ is $\theta_1-\theta_2-\theta_3-\theta_4=\pi/2$ and ${\hat P}_\mu \cdot {\hat P}_\nu =-1/3$ ($\mu \neq \nu$), suggestive of a nonzero value of the chirality $\chi_{ijk}={\bf S}_i \cdot ({\bf S}_j \times {\bf S}_k)$ formed by any three spins on each tetrahedron. It will be shown in the following numerical calculations that this quadruple-${\bf q}$ $(\frac{1}{2}, \frac{1}{2}, \frac{1}{2})$ state with the nonzero chirality is nothing but the hedgehog lattice. It should be emphasized that at least at the MF level, an infinitesimally small breathing alternation yields nonzero $\varepsilon$ and resultantly, can induce the $(\frac{1}{2}, \frac{1}{2}, \frac{1}{2})$ hedgehog lattice.

\section{Result of the MC simulation at zero field}
Now that the role of the breathing bond-alternation has been understood, we shall perform a rigorous analysis of the finite-temperature properties of the Hamiltonian (\ref{eq:Hamiltonian}) by means of Monte Carlo (MC) simulations. In this section, we consider the zero-field case of $H=0$.

In our MC simulations, 2$\times 10^5$ sweeps are carried out under the periodic boundary condition and the first half is discarded for thermalization, where our 1 MC sweep consists of 1 heatbath sweep and successive 10 overrelaxation sweeps. Observations are done at every MC sweep and the statistical average is taken over 4 independent runs starting from different random initial configurations. We identified the low-temperature ordered phase by observing various physical quantities such as the spin collinearity $P = \frac{3}{2} \big\langle \frac{1}{N^2}\sum_{i,j} \big( {\bf S}_i\cdot{\bf S}_j\big)^2 - \frac{1}{3} \big\rangle$ and the spin and chirality structure factors $F_{S}({\bf q}) = \big\langle \big| \frac{1}{N} \sum_i  {\bf S}_i \, e^{i{\bf q}\cdot{\bf r}_i}\big|^2\big\rangle$ and $F_{C}({\bf q}) = \big\langle \big| \frac{1}{N/4} \sum_l  \chi({\bf R}_l) \, e^{i{\bf q}\cdot{\bf R}_l}\big|^2\big\rangle$, where the scalar chirality of the $l$th tetrahedron with its center-of-mass position ${\bf R}_l$ is defined by $\chi({\bf R}_l) = \sum_{i,j,k \in l \, {\rm th} \, {\rm tetra}} \chi_{ijk}$ and the order of $i$, $j$, and $k$ is defined in the anticlockwise direction with respect to the normal vector of the triangle formed by the three sites $i$, $j$, and $k$, $\hat{n}_{ijk}$, pointing outward from ${\bf R}_l$. In the same manner, we define the solid angle subtended by the four spins on the tetrahedron as $\Omega({\bf R}_l)=\sum_{i,j,k \in l \, {\rm th} \, {\rm tetra}}\Omega_{ijk}$. 

The temperature dependence of the specific heat $C$ and the spin collinearity $P$ obtained in the MC simulations for $J_2/J_1=0$ and $J_3/J_1=0.7$ is shown in Fig. \ref{fig:hedgehog} (a). In the uniform pyrochlore lattice with $J'_1/J_1=1$, $P$ starts developing at the transition temperature indicated by the sharp peak in $C$, and approaches the maximum value of 1 at the lowest temperature. Since the Hamiltonian (\ref{eq:Hamiltonian}) itself does not have an interaction to collinearize spins, thermal fluctuations should be relevant to the spin collinearity as is often the case with frustrated magnets \cite{fcc_Henly_87, tri_Kawamura_85}. With increasing the breathing bond-alternation, i.e., decreasing $J'_1/J_1$, such a temperature effect is gradually suppressed. At $J'_1/J_1=0.6$, $P$ develops on cooling similarly to the case of $J'_1/J_1=1$, but it finally goes to zero at $T=0$ via an intermediate phase explained below, suggesting the occurrence of a noncollinear ground state. At the smaller value of $J'_1/J_1=0.2$, the ordered phase is completely masked by the noncollinear state with $P=0$. To look into the details of the ordered phases, we examine the spin and chirality structure factors $F_{S}({\bf q})$ and $F_{C}({\bf q})$. 

Figure \ref{fig:hedgehog} (b) shows $F_{S}({\bf q})$ in the noncollinear state with $P=0$. This state is characterized by the magnetic Bragg peaks with the same intensity at the cubic-symmetric families of $(\pm\frac{1}{2}, \pm\frac{1}{2}, \pm\frac{1}{2})$, so that it is a cubic-symmetric quadruple-${\bf q}$ state. Such a quadruple-$(\frac{1}{2}, \frac{1}{2}, \frac{1}{2})$ spin structure can commonly be seen irrespective of the value of $J'_1/J_1$. One can see from Fig. \ref{fig:hedgehog} (a) that the averaged intensity of the Bragg peaks $O^S_{(\frac{1}{2}\frac{1}{2}\frac{1}{2})} = 2 \sum_{h,k,l=\pm 1/2} F_S (h,k,l)$ serves as the order parameter to characterize the transition from the paramagnetic phase in all the three cases of $J'_1/J_1=1$, $0.6$, and $0.2$. Although whether the spins are collinear or noncollinear is indistinguishable in the spin sector $F_{S}({\bf q})$, it is clearly visible in the chirality sector $F_{C}({\bf q})$. In the uniform case of $J'_1/J_1=1$, spins are collinearly ordered, so that the local scalar chirality $\chi_{ijk}$ and the associated $F_{C}({\bf q})$ are vanishingly small. In contrast, in the breathing case of $J'_1/J_1<1$, $F_{C}({\bf q})$ in the noncollinear state with $P=0$ exhibits Bragg peaks at $(\pm\frac{1}{2}, \pm\frac{1}{2}, \pm\frac{1}{2})$ suggestive of a tetrahedron-based chirality order [see Fig. \ref{fig:hedgehog} (c)]. The averaged Bragg intensities in the chirality sector $O^C_{(\frac{1}{2}\frac{1}{2}\frac{1}{2})} = \frac{1}{4}\sum_{h,k,l=\pm 1/2} F_C (h,k,l)$ for $J'_1/J_1=1$, $0.6$, and $0.2$ are shown in the bottom panel of Fig. \ref{fig:hedgehog} (a). One can see that $O^C_{(\frac{1}{2}\frac{1}{2}\frac{1}{2})}$ serves as the order parameter for the noncollinear $(\frac{1}{2},\frac{1}{2},\frac{1}{2})$ state with $P=0$. 

\begin{figure}[t]
\begin{center}
\includegraphics[scale=0.75]{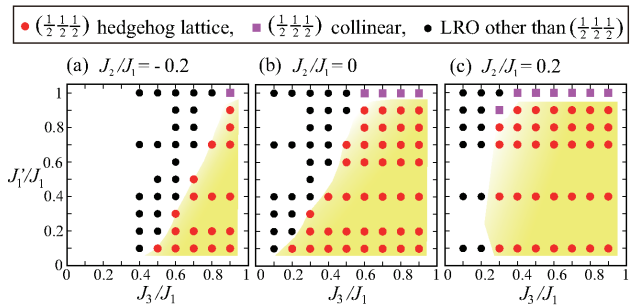}
\caption{The parameter dependence of the low-temperature ordered phases at $H=0$. 
(a) $J_2/J_1=-0.2$, (b) $J_2/J_1=0$, and (c) $J_2/J_1=0.2$.
The stability regions are determined from the MC results at $T/J_1=0.01$. The $(\frac{1}{2},\frac{1}{2},\frac{1}{2})$ hedgehog-lattice and collinear phases are obtained at red and pink colored points, respectively, whereas LRO's with ordering vectors other than $(\frac{1}{2},\frac{1}{2},\frac{1}{2})$ are obtained at black colored points. 
The stability regions are consistent with the result of the $J^{\alpha\beta}({\bf q})$ analysis (see Fig. \ref{fig:Jq_Paradep} in Appendix A) \label{fig:Jdep}}
\end{center}
\end{figure}
The real-space spin configuration of the noncollinear state is shown in Fig. \ref{fig:hedgehog_snapshot}. Spins belonging to each sublattice, which are represented by the same colored arrows in Fig. \ref{fig:hedgehog_snapshot}, constitute up-down antiferromagnetic chains along the bond directions, and four sublattices have different polarization axes. In Fig. \ref{fig:hedgehog_snapshot}, since four outgoing (incoming) spins on a tetrahedron subtend the solid angle of $\Omega({\bf R}_l)=4\pi$ ($-4\pi$), the upper (lower) small tetrahedron at the center of the cubic unit cell has the topological charge of $+1$ ($-1$). As the spins are ordered in a way such that the two cubic unit cells are alternately arranged, the state is nothing but the hedgehog lattice in which the same number of the monopoles with $+1$ charge and the antimonopoles with $-1$ charge are equally populated and their densities $n_+$ and $n_-$ are given by $n_+=n_-=\frac{1}{16}$.
For the monopole and antimonopole tetrahedra, the total spin on each tetrahedron is zero as is expected from the NN antiferromagnetic interaction $J_1$ or $J'_1$ \cite{Reimers_MF}, while not for other tetrahedra. This is because the $(\frac{1}{2}, \frac{1}{2}, \frac{1}{2})$ spin ordering in the present system is driven by the relatively strong third NN antiferromagnetic interaction $J_3$ rather than $J_1$ or $J'_1$ and thus, not all the tetrahedra satisfy the constraint of total spin zero stemming from $J_1$ and $J'_1$. In relation to this, as one can see from Fig. \ref{fig:hedgehog_snapshot}, the monopoles and antimonopoles, which satisfy the constraint, are formed not on the large tetrahedra but on the small tetrahedra, since the latter has the stronger NN antiferromagnetic interaction $J_1>J'_1$ and thus, tries to follow the constraint as much as possible.
We note that in the present system, the isotropic Heisenberg spins spontaneously form this chirality order as a result of the frustration, which is in sharp contrast to spin-ice systems with strong local magnetic anisotropy where a similar $(\frac{1}{2}, \frac{1}{2}, \frac{1}{2})$ spin structure has been discussed \cite{Ice_Ishizuka_11}.

In the quadruple-${\bf q}$ $(\frac{1}{2}, \frac{1}{2}, \frac{1}{2})$ phases other than the hedgehog lattice, relative angles between the polarization axes for the four sublattices are changed with each up-down antiferromagnetic chain being almost unchanged. Figure \ref{fig:hedgehog} (d) shows the map of the spins in the three different $(\frac{1}{2}, \frac{1}{2}, \frac{1}{2})$ phases obtained at $J_1'/J_1=0.6$. One can see from the bottom (top) panel of Fig. \ref{fig:hedgehog} (d) that in the higher-temperature collinear (lower-temperature hedgehog-lattice) phase, the polarization axes for the four sublattices are the same (different). The spin ordering pattern in the collinear phase is up-up-down-down along all the bond directions \cite{Site_AK_16, Site_AK_19}. As shown in the middle of Fig. \ref{fig:hedgehog} (d), the intermediate-temperature phase between the two is a coplanar state in which four sublattices are divided into two pairs and spins belonging to each pair have the same polarization axis (see Fig. \ref{fig:snap_comp} in Appendix B for details of the real-space spin configurations).

The stability region of the $(\frac{1}{2}, \frac{1}{2}, \frac{1}{2})$ hedgehog lattice is summarized in Fig. \ref{fig:Jdep}. Among the three quadruple-${\bf q}$ $(\frac{1}{2}, \frac{1}{2}, \frac{1}{2})$ states, only the coplanar one cannot survive down to the lowest temperature, so that it does not appear in Fig. \ref{fig:Jdep}. The $(\frac{1}{2}, \frac{1}{2}, \frac{1}{2})$ hedgehog lattice is realized in the wide range of the parameter space even in the presence of $J_2$.
As expected from the MF analysis, a small breathing alternation, i.e., small deviation $1- J'_1/J_1 $, seems to be sufficient to stabilize the hedgehog lattice.

\section{Effect of a magnetic field}
\begin{figure}[t]
\begin{center}
\includegraphics[scale=0.48]{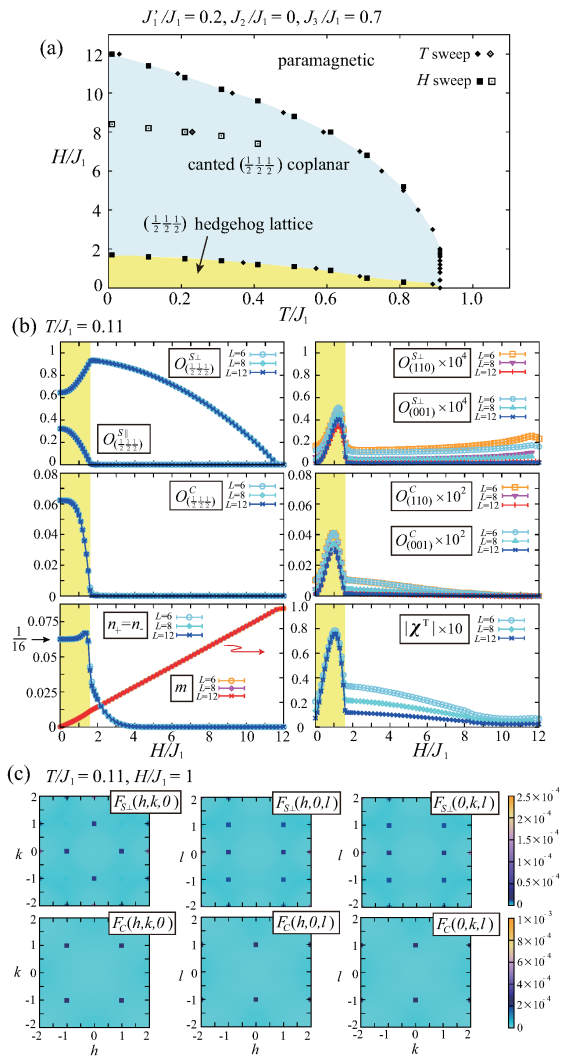}
\caption{MC results obtained for $J'_1/J_1=0.2$, $J_2/J_1=0$, and $J_3/J_1=0.7$. (a) The temperature and magnetic-field phase diagram. (b) Field dependence of various quantities at $T/J_1=0.11$. Top (center) panels: the average spin (chirality) Bragg intensities $O^{S_\parallel}_{(\frac{1}{2}\frac{1}{2}\frac{1}{2})}$, $O^{S_\perp}_{(\frac{1}{2}\frac{1}{2}\frac{1}{2})}$ [$O^{C}_{(\frac{1}{2}\frac{1}{2}\frac{1}{2})}$], $O^{S_\perp}_{(001)}$ [$O^{C}_{(001)}$], and $O^{S_\perp}_{(110)}$ [$O^{C}_{(110)}$], where $S_\parallel$ and $S_\perp$ denote spin components parallel and perpendicular to the field, respectively. Bottom left panel: the monopole (or equivalently, antimonopole) density $n_+$ ($=n_-$) and the magnetization $m$. Bottom right panel: the total spin chirality responsible for the topological Hall effect $|\mbox{\boldmath $\chi$}^{\rm T}|$. (c) Spin and chirality structure factors $F_{S_\perp}({\bf q})$ and $F_C({\bf q})$ in the $(h,k,0)$ (left), $(h,0,l)$ (center), and $(0,k,l)$ (right) planes obtained at $T/J_1=0.11$ and $H/J_1=1$. 
\label{fig:HT}}
\end{center}
\end{figure}
In this section, we will discuss the effect of a magnetic field $H$ on the $(\frac{1}{2}, \frac{1}{2}, \frac{1}{2})$ hedgehog lattice.
Figure \ref{fig:HT} (a) shows the temperature and magnetic-field phase diagram for $J'_1/J_1=0.2$, $J_2/J_1=0$, and $J_3/J_1=0.7$. The low-field and high-field phases are the $(\frac{1}{2}, \frac{1}{2}, \frac{1}{2})$ hedgehog-lattice and coplanar states being subject to spin canting, respectively. We note that for larger values of $J'_1/J_1$, the canted collinear state also shows up (see Fig. \ref{fig:HT_J06} in Appendix C). Although a weak transition-like anomaly is found in the high-field phase [see open symbols in Fig. \ref{fig:HT} (a)], the detailed analysis of the high-field phase is beyond the scope of this work, and hereafter, we will focus on the low-field hedgehog-lattice phase. As shown in the left panels of Fig. \ref{fig:HT} (b), the in-field hedgehog lattice is characterized by the enhanced (suppressed) $(\pm\frac{1}{2}, \pm\frac{1}{2}, \pm\frac{1}{2})$ magnetic Bragg reflections for the spin component parallel (perpendicular) to the field $O^{S_\parallel}_{(\frac{1}{2}\frac{1}{2}\frac{1}{2})}$ \big[$O^{S_\perp}_{(\frac{1}{2}\frac{1}{2}\frac{1}{2})}$\big] as well as the nonzero chirality order parameter $O^{C}_{(\frac{1}{2}\frac{1}{2}\frac{1}{2})}$ and the monopole or antimonopole density of $n_+=n_-=\frac{1}{16}$. Also, the magnetization $m$ shows a weak downward convex curve as a function of $H$ in this low-field phase. Besides, in a magnetic field, additional weak Bragg peaks emerge at the wave vectors of the $(0,0,1)$ and $(1,1,0)$ families in both $F_S({\bf q})$ and $F_C({\bf q})$, which can clearly be seen in Fig. \ref{fig:HT} (c). This in-field hedgehog lattice is tetragonal symmetric in the sense that among the three cubic symmetric points of $(\pm 1,0,0)$, $(0,\pm 1,0)$, and $(0,0,\pm 1)$, only one is selected, and such a situation is also the case for the $(1,1,0)$ family. In the case of Fig. \ref{fig:HT} (c), $z$-direction is special. In the chirality sector [the lower panels of Fig. \ref{fig:HT} (c)], only $(0,0,1)$ and $(1,1,0)$ are picked up, whereas in the spin sector (the upper ones), the rest four, i.e., $(1,0,0)$, $(0,1,0)$, $(1,0,1)$ and $(0,1,1)$, are selected. As one can see from the right panels of Fig. \ref{fig:HT} (b), the averaged intensities of these additional peaks in the spin and chirality sectors $O^{S_\perp}_{(001)}$, $O^{S_\perp}_{(110)}$, $O^{C}_{(001)}$, and $O^{C}_{(110)}$ are nonzero only in the in-field hedgehog lattice. 

When the system is metallic and localized spins ${\bf S}_i$ are coupled to conduction electrons, the field-induced symmetry reduction from cubic to tetragonal in the hedgehog lattice is reflected in the Hall effect of spin chirality origin, i.e., the topological Hall effect. In the weak-coupling theory \cite{THE_Tatara_02}, the topological Hall conductivity $\sigma^{\rm T}_{\mu\nu}$ is proportional to the total spin chirality multiplied by a geometrical factor $\mbox{\boldmath $\chi$}^{\rm T} =\sum_{\langle i,j,k \rangle_{S,L}} \chi_{ijk} \, \hat{n}_{ijk}$, namely, $\sigma^{\rm T}_{\mu\nu} \propto \epsilon_{\mu\nu\rho} \chi^{\rm T}_\rho$, where ${\langle i,j,k \rangle}_{S,L}$ denotes the summation over all the three sites on each tetrahedron and $\hat{n}_{ijk}$ represents the surface normal. $\mbox{\boldmath $\chi$}^{\rm T}$ corresponds to the emergent fictitious magnetic field. Although according to Ref. \cite{THE_Tatara_02}, $\mbox{\boldmath $\chi$}^{\rm T}$ should be calculated by taking further NN sites into account, we have restricted only to the NN triads because the dominant contribution comes from such short-distance triads. 
Indeed, we have checked that the original and present $\mbox{\boldmath $\chi$}^{\rm T}$'s exhibit qualitatively the same field dependence. 
One can see from the bottom right panel of Fig. \ref{fig:HT} (b) that with increasing the external magnetic field, the fictitious field $|\mbox{\boldmath $\chi$}^{\rm T}|$ is gradually induced in the hedgehog-lattice phase and it vanishes out of this topological phase. It is also found that the direction of $\mbox{\boldmath $\chi$}^{\rm T}$ is not arbitrary but rather takes one of the six degenerate directions, $\pm \hat{x}$, $\pm \hat{y}$, and $\pm \hat{z}$. As inferred from the similar field dependences of $O^{S_\perp}_{(001)}$, $O^{S_\perp}_{(110)}$, $O^{C}_{(001)}$, $O^{C}_{(110)}$, and $|\mbox{\boldmath $\chi$}^{\rm T}|$ [see right panels of Fig. \ref{fig:HT} (b)], the preferred direction of $\mbox{\boldmath $\chi$}^{\rm T}$ is determined from the tetragonal symmetry of the spin structure. For example, when the $z$-direction is picked up for the tetragonal spin structure like in the case of Fig. \ref{fig:HT} (c), $\mbox{\boldmath $\chi$}^{\rm T}$ is along the $\hat{z}$-direction, so that $\sigma^{\rm T}_{xy}=-\sigma^{\rm T}_{yx}$ takes a finite value while other components are zero.
In the next section, we will address the origin of this unique type of the topological Hall effect characteristic of the $(\frac{1}{2},\frac{1}{2},\frac{1}{2})$ hedgehog lattice.

\section{Origin of the field-induced topological Hall effect}
\begin{figure}[t]
\begin{center}
\includegraphics[scale=0.4]{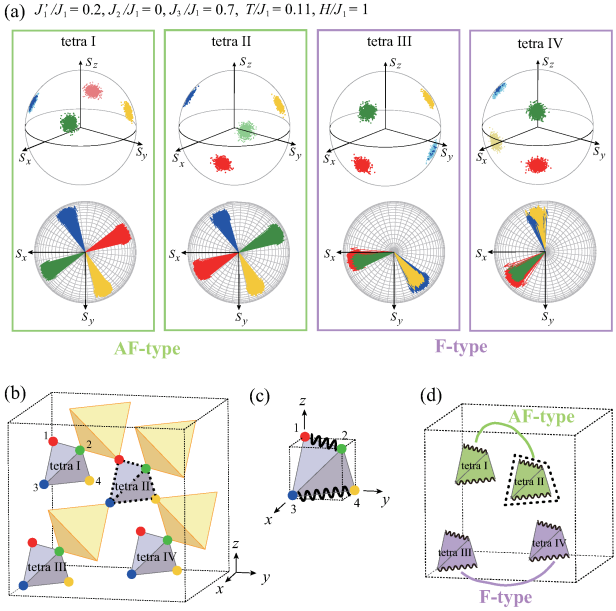}
\caption{Tetragonal-symmetric spin structure in the in-field hedgehog-lattice phase. (a) MC snapshots of spins mapped onto a unit sphere (upper panels) and their projection onto the $S_xS_y$ spin plane (lower panels) for the same parameters as those in Fig. \ref{fig:HT} (c), where tetra's I, II, III, and IV represent the four tetrahedra in the cubic unit cell shown in (b) and the color notations are the same as those in Figs. \ref{fig:hedgehog_snapshot} and \ref{fig:hedgehog}. The spin configurations of tetra's I and II (III and IV) are classified as AF-type (F-type) (see the main text). (c) Tetragonal-symmetric pairing pattern in each tetrahedron, where wavy lines denote pair bonds. (d) Tetragonal-symmetric distribution of the AF-type (green) and F-type (purple) tetrahedra within the cubic unit cell. In (b) and (d), a monopole tetrahedron is outlined by black dots. In (a), short-time average over 10 MC steps has been made to reduce the thermal noise. \label{fig:tetragonalspin}}
\end{center}
\end{figure}
To examine the association between the field-induced tetragonal-symmetric hedgehog-lattice and the net chirality $\mbox{\boldmath $\chi$}^{\rm T}$, or equivalently, the emergent fictitious field for the topological Hall effect, we shall first discuss the magnetic structure of the in-field hedgehog lattice. Figure \ref{fig:tetragonalspin} (a) shows the spin configuration associated with Fig. \ref{fig:HT} (c). Since the $(\frac{1}{2}, \frac{1}{2}, \frac{1}{2})$ hedgehog lattice consists of the alternating array of two cubic unit cells having oppositely oriented spins [see Fig. \ref{fig:hedgehog_snapshot}], only one of the two is shown in Fig. \ref{fig:tetragonalspin}. One can see from the lower panels of Fig. \ref{fig:tetragonalspin} (a) that the sublattices 1 and 2, and 3 and 4 are paired up and two spins in each pair are collinearly aligned in the $S_xS_y$ spin plane. Due to this pairing, the symmetry of each tetrahedron is reduced to tetragonal with respect to the $z$-axis [see Fig, \ref{fig:tetragonalspin} (c)]. Possible three pairing patterns, i.e, stacking-directions of the pair bonds, correspond to possible three axes for the tetragonal spin structure.

\begin{figure*}[t]
\begin{center}
\includegraphics[scale=0.48]{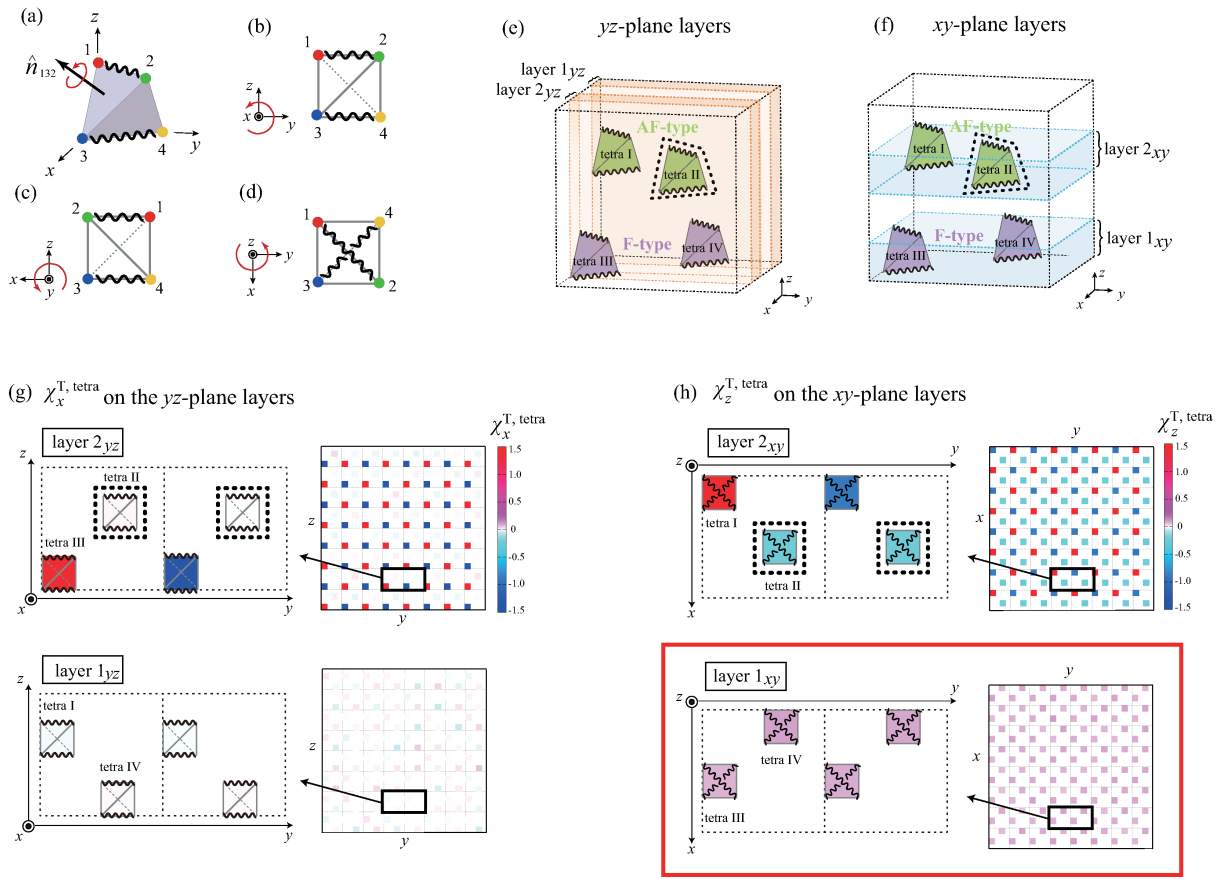}
\caption{Spin and chirality configurations on the projected two-dimensional planes. (a) A tetrahedron having a tetragonal-symmetric pairing with respect to the $z$-axis, and (b)-(d) its projections onto the two-dimensional planes perpendicular to $x$-, $y$-, and $z$-axes, respectively, where a red circular arrow denotes the direction in which the spin chirality $\chi_{ijk}$ is defined and other notations are the same as those in Fig. \ref{fig:tetragonalspin}. (e) and (f) Layers of one-small-tetrahedron width defined for the tetragonal-symmetric spin structure in Fig. \ref{fig:tetragonalspin} (d). (g) and (h) Layer-resolved distributions of the chirality calculated from the MC snapshot in Fig. \ref{fig:tetragonalspin}. In (g) [(h)], $\chi_x^{\rm T, tetra}$ ($\chi_z^{\rm T, tetra}$) on the $yz$- ($xy$-)plane layers defined in (e) [(f)] is shown, where the left and right tetrahedra outlined by black dots correspond to the monopole and antimonopole, respectively. The $xy$-plane layer without the monopoles, which is enclosed by a red box in (h), yields a dominant contribution to the total chirality. In (g) and (h), short-time average over 200 MC steps has been made to reduce the thermal noise.   
\label{fig:schi}}
\end{center}
\end{figure*}
Once a pairing symmetry is fixed, there exist two types of the tetrahedral spin configurations, which we call AF- and F-types. As shown in Fig. \ref{fig:tetragonalspin} (a), in the AF-type (F-type) tatrahedron, the $S_xS_y$ components in each pair are antiferromagnetically (ferromagnetically) aligned, whereas the $S_z$ ones ferromagnetically (antiferromagnetically). Also, the collinear axes for the two pairs are orthogonal to each other in the AF-type, while not in the F-type. The AF-type is further categorized into two: one involves $S_z$'s of the same sign [tetra I in Fig. \ref{fig:tetragonalspin} (a)] and the other involves $S_z$'s of opposite signs [tetra II in Fig. \ref{fig:tetragonalspin} (a)]. The monopole and antimonopole correspond to the latter. The distribution of these AF- and F-type tetrahedra within the cubic unit cell is also tetragonal-symmetric. One can see from Fig. \ref{fig:tetragonalspin} (d) that the AF-type tetrahedron pair and the F-type one are stacking along the tetragonal-symmetric $z$-axis. Thus, the symmetry-reduction to tetragonal occurs at the level of the cubic unit cell as well as each tetrahedron. 

Now that the spin structure of the in-field hedgehog lattice is clarified, we shall next discuss how it is related to the total spin chirality $\mbox{\boldmath $\chi$}^{\rm T}=\sum_{\langle i,j,k \rangle_{S,L}} \chi_{ijk} \, \hat{n}_{ijk}$. As a first step, we consider the associated local chirality $\mbox{\boldmath $\chi$}^{\rm T,tetra}=\sum_{i,j,k \in {\rm tetra}} \chi_{ijk} \, \hat{n}_{ijk}$ for the small tetrahedron shown in Fig. \ref{fig:schi} (a), which is calculated as
\begin{equation}\label{eq:tetra_schi}
\mbox{\boldmath $\chi$}^{\rm T,tetra} = \frac{1}{\sqrt{3}}\left( \begin{array}{c}
\chi_{132}+\chi_{234}-\chi_{124}-\chi_{143} \\
-\chi_{132}+\chi_{234}+\chi_{124}-\chi_{143}\\
\chi_{132}-\chi_{234}+\chi_{124}-\chi_{143} 
\end{array} \right) .
\end{equation}  
When the tetrahedron is projected onto the two-dimensional planes perpendicular to the $x$-, $y$-, and $z$-axes [see Figs. \ref{fig:schi} (b)-(d)], the chirality defined on each two-dimensional plane $\chi^{\rm 2D,tetra}$ is given by a summation of four different $\chi_{ijk}$'s with triad $(i,j,k)$ ordered in the anticlockwise direction with respect to the surface normal. Then, one finds that the so-obtained $\chi^{\rm 2D,tetra}$'s correspond to the $x$-, $y$-, and $z$-components of $\mbox{\boldmath $\chi$}^{\rm T,tetra}$ in Eq. (\ref{eq:tetra_schi}), which suggests that the $\rho$ component of $\mbox{\boldmath $\chi$}^{\rm T}$ is determined from the spin configurations on the projected two-dimensional planes perpendicular to the $\rho$-direction. Thus, we next consider how the tetragonal symmetry in each tetrahedron, i.e., the pairing pattern, looks on the projected planes. 

One can see from wavy lines in Figs. \ref{fig:schi} (a)-(d) that the paired spins are arranged diagonally on the plane perpendicular to the tetragonal-symmetric direction ($z$-direction in Fig. \ref{fig:schi}), while adjacently on other planes. Such a difference in the pairing arrangement yields an anisotropy in the local chirality. 
It turns out from a simple calculation (see Appendix D) that $\chi^{\rm 2D,tetra}$ on the projected plane becomes nonzero for the diagonal arrangement, while it is basically zero for the adjacent ones except a particular case of the F-type spin configuration, which corresponds to Fig. \ref{fig:schi_isotetra} (f) in Appendix D.
Bearing the association between the tetragonal symmetry and the {\it local} chirality in our mind, we will turn to the {\it total} chirality on the projected two-dimensional planes, i.e. layers of one-tetrahedron width [see Figs. \ref{fig:schi} (e) and (f)].

Figures \ref{fig:schi} (g) and (h) show the layer-resolved distributions of $\mbox{\boldmath $\chi$}^{\rm T,tetra}$ calculated from the spin configuration in Fig. \ref{fig:tetragonalspin}. 
One can see from Fig. \ref{fig:schi} (g) that on the layers perpendicular to the non-tetragonal-symmetric $x$-direction where the paired spins are arranged adjacently, the local chirality $\chi_x^{\rm T,tetra}$ for most of the tetrahedra takes a vanishingly small value, which may be positive or negative depending on the thermal noise. Although an exceptional F-type tetrahedron, e.g., tetra III in the zoomed view in Fig. \ref{fig:schi} (g), yields a nonzero large $\chi_x^{\rm T,tetra}$, it is completely canceled out by the contribution from the counter tetrahedron in the neighboring cubic unit cell, so that the net chirality vanishes on any $yz$-plane layer. Such a situation is also the case for the layers perpendicular to the $y$-direction.

By contrast, on the layers perpendicular to the tetragonal-symmetric $z$-direction where the paired spins are arranged diagonally, not only the local chirality $\chi_z^{\rm T,tetra}$ but also the net chirality becomes nonzero.
As shown in Fig. \ref{fig:schi} (h), $\chi_z^{\rm T,tetra}$'s on the layer without monopoles [layer $1_{xy}$ in Fig. \ref{fig:schi} (h)] take nonzero values of the same sign, leading to a large total chirality, while $\chi_z^{\rm T,tetra}$'s on the layer with monopoles [layer $2_{xy}$ in Fig. \ref{fig:schi} (h)] take positive and negative signs. As one can see from the zoomed view focusing on the two cubic unit cells in Fig. \ref{fig:schi} (h), the monopole and antimonopole tetrahedra yield $\chi_z^{\rm T,tetra}$'s of the same sign and their contributions are canceled out by those from other tetrahedra.
Thus, a dominant contribution comes from the layers {\it without} the monopoles, which has actually been confirmed from the numerical data for Fig. \ref{fig:schi} (h) together with the result of analytical calculations shown in Appendix D.

\section{Summary and Discussion}
In this paper, we have demonstrated by means of MC simulations that the hedgehog lattice characterized by the $(\frac{1}{2},\frac{1}{2},\frac{1}{2})$ magnetic Bragg reflections is induced by the frustration in classical Heisenberg antiferromagnets on the breathing pyrochlore lattice, where the breathing bond-alternation quantified by the ratio of the NN antiferromagnetic interactions $J_1'/J_1$ and a large third NN one along the bond directions $J_3$ are essential. It is also found that in a magnetic field, the structure of the $(\frac{1}{2},\frac{1}{2},\frac{1}{2})$ hedgehog lattice is changed from cubic to tetragonal, resulting in the nonzero net spin chirality $\mbox{\boldmath $\chi$}^{\rm T}$, or equivalently, nonzero emergent fictitious magnetic field, along the tetragonal-symmetric direction.  

In the DM-induced hedgehog-lattice phase, the fictitious field is also caused by the magnetic field, but its origin is the field-induced position shifts of monopoles and antimonopoles \cite{Hedgehog_MFtheory_Park_11}, which is in sharp contrast to the present frustrated system where their positions are unchanged. As a result, the total spin chirality $\mbox{\boldmath $\chi$}^{\rm T}$ appears in any of the possible three directions $x$, $y$, and $z$ in the present system, while only along the applied field direction in the DM system. 
In addition, the sign of $\mbox{\boldmath $\chi$}^{\rm T}$ can be positive or negative in the former, while it is fixed by the DM interaction in the latter. Such a degeneracy of the right-handed and left-handed chiralities is a unique aspect of the frustration-induced topological spin textures distinct from the DM-induced ones \cite{SkX_Okubo_12}.

In experiments, although various classes of breathing pyrochlore antiferromagnets such as the chromium oxides Li(Ga, In)Cr$_4$O$_8$ \cite{BrPyro_Okamoto_13, BrPyro_Tanaka_14,BrPyro_Nilsen_15,BrPyro_Saha_16,BrPyro_Lee_16,BrPyro_Saha_17,BrPyro_doped_Okamoto_15,BrPyro_doped_Wang_17,BrPyro_doped_Wawrzynczak_17,BrPyro_Hdep_Okamoto_17,BrPyro_Hdep_Gen_19} and sulfides Li(Ga, In)Cr$_4$S$_8$ \cite{BrPyro_Sulfides_Okamoto_18, BrPyro_Sulfides_Pokharel_18,BrPyro_Hdep_Gen_20} and the quantum magnet Ba$_3$Yb$_2$Zn$_5$O$_{11}$ \cite{qBrPyro_Kimura_14, qBrPyro_Haku_prb16, qBrPyro_Haku_jpsj16, qBrPyro_Rau_16,qBrPyro_Rau_18} have been studied, the $(\frac{1}{2},\frac{1}{2},\frac{1}{2})$ spin correlation has not been reported so far. On the other hand, the uniform pyrochlore antiferromagnets Ge$B_2$O$_4$ ($B$=Ni, Co, Fe, Cu) \cite{GeNiO_Crawford_03,GeNiCoO_Diaz_06,GeNiO_Lancaster_06,GeNiO_Matsuda_08,GeCoO_Watanabe_08,GeCoO_Watanabe_11,GeCoO_Tomiyasu_11,GeAO_Barton_14,GeCoO_Fabreges_17,GeCoO_Pramanik_19,GeCuO_Zou_16} do exhibit $(\frac{1}{2},\frac{1}{2},\frac{1}{2})$ magnetic LRO, although the experimentally proposed spin structure as well as the estimated exchange interactions (e.g., ferromagnetic $J_1$) seems to be different from the ones in the present theoretical model. If one can introduce the breathing alternation in the Ge$B_2$O$_4$ family, modifying the exchange interactions, the $(\frac{1}{2},\frac{1}{2},\frac{1}{2})$ hedgehog lattice might be realized. Although the above pyrochlore magnets are insulators, when the system is tuned to be metallic, the hedgehog lattice, if it is realized, should be evidenced distinctly by the topological Hall effect associated with the total spin chirality $\mbox{\boldmath $\chi$}^{\rm T}$ together with the $(\frac{1}{2},\frac{1}{2},\frac{1}{2})$ magnetic Bragg reflections. 

We believe that this work presenting the novel mechanism different from the conventional DM interaction to stabilize the hedgehog lattice will promote the exploration of new classes of magnets hosting topological spin textures.

\begin{acknowledgments}
The authors thank K. Tomiyasu and S. Ishiwata for useful discussions. We are thankful to ISSP, the University of Tokyo and YITP, Kyoto University for providing us with CPU time. This work is supported by JSPS KAKENHI Grant Number JP17H06137.
\end{acknowledgments}

\appendix
\section{Ordering wave-vector with the lowest eigen value of $J^{\alpha\beta}({\bf q})$}
\begin{figure}[t]
\begin{center}
\includegraphics[scale=0.75]{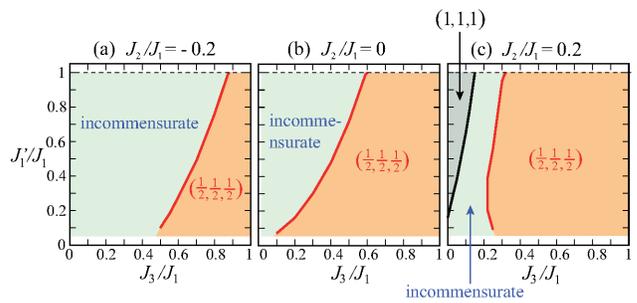}
\caption{Ordering wave-vectors having the lowest eigen value of $J^{\alpha\beta}({\bf q})$ defined in Eq. (\ref{eq:H_Ising_FT}). (a) $J_2/J_1=-0.2$, (b) $J_2/J_1=0$, and (c) $J_2/J_1=0.2$. The commensurate wave-vectors $(\frac{1}{2}, \frac{1}{2}, \frac{1}{2})$ and $(1,1,1)$ are represented in units of $\frac{2\pi}{a}$, where $a$ is a side length of the cubic unit cell. \label{fig:Jq_Paradep}}
\end{center}
\end{figure}

In general, an ordering wave-vector minimizing the eigen value of $J^{\alpha\beta}({\bf q})$ defined in Eq. (\ref{eq:H_Ising_FT}) should be realized in the ground state, provided that the local spin-length constraints $|{\bf S}_i|=1$ are satisfied. Thus, searching the ordering wave-vector with the lowest eigen value is a starting point to analyze ordering properties of the system. 
Figure \ref{fig:Jq_Paradep} shows ordering wave-vectors with the lowest eigen value of $J^{\alpha\beta}({\bf q})$. For larger values of $J_3/J_1$, the ground-state candidate is a state with the ordering wave-vector of $(\pm\frac{1}{2}, \pm\frac{1}{2}, \pm\frac{1}{2})$, whereas for smaller values of $J_3/J_1$, it is a state with an incommensurate wave-vector. With increasing $J_2/J_1$ or decreasing $J_1'/J_1$, the $(\frac{1}{2}, \frac{1}{2}, \frac{1}{2})$ state gets more stable. A commensurate $(1,1,1)$ state is also possible for an antiferromagnetic $J_2$ and a weak $J_3$.   \\

\section{Real-space spin configurations of the three different quadruple-{\bf q} $(\frac{1}{2},\frac{1}{2},\frac{1}{2})$ states}
\begin{figure}[t]
\begin{center}
\includegraphics[scale=0.4]{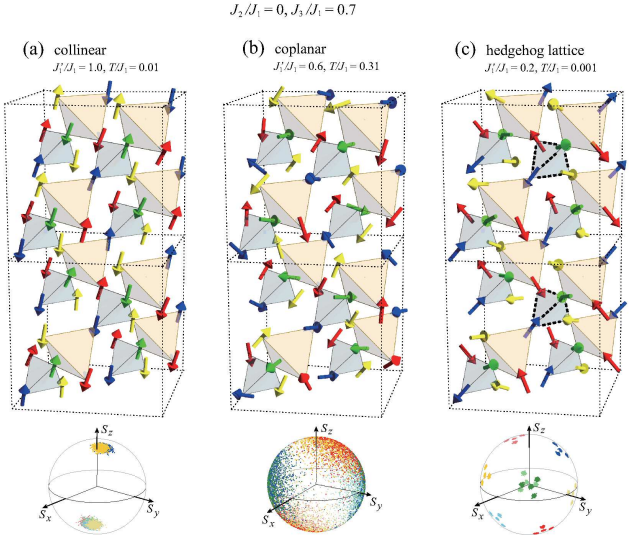}
\caption{MC snapshots taken in the three different quadruple-{\bf q} $(\frac{1}{2},\frac{1}{2},\frac{1}{2})$ states for $J_2/J_1=0$, $J_3/J_1=0.7$, and $H=0$. (a) the collinear state at $J_1'/J_1=1$ and $T/J_1=0.01$, (b) the coplanar state at $J_1'/J_1=0.6$ and $T/J_1=0.31$, and (c) the hedgehog-lattice state at $J_1'/J_1=0.2$ and $T/J_1=0.001$. In each figure, the upper panel shows the real-space spin configuration, and the lower one shows the associated all the spins mapped onto a unit sphere. The lower panel of (b) is exactly the same as the middle panel of Fig. \ref{fig:hedgehog} (d). The notations of symbols and colors are the same as those of Figs. \ref{fig:hedgehog_snapshot} and \ref{fig:hedgehog}. \label{fig:snap_comp}}
\end{center}
\end{figure}

In the MC simulations, we obtain the three different quadruple-{\bf q} $(\frac{1}{2},\frac{1}{2},\frac{1}{2})$ states, the collinear, coplanar, and hedgehog-lattice ones. Their real-space spin configurations at $H=0$ are shown in Fig. \ref{fig:snap_comp}. In all the three cases, spins belonging to each sublattice (the same colored arrows in Fig. \ref{fig:snap_comp}) constitute up-down antiferromagnetic chains along the bond directions, although exactly speaking, the spin directions in each chain are not perfectly collinear when the lattice is not uniform, which will be explained below. 

The collinear phase can survive down to a low temperature basically for the uniform pyrochlore lattice with $J_1'/J_1 = 1$. Since the polarization axes of the different sublattices are collinearly aligned, the resultant spin configuration consists of the up-up-down-down antiferromagnetic chains running along all the bond directions. In this case, the spin directions in each chain are perfectly collinear.

The coplanar phase is realized only at moderate temperatures. Indeed, when we evaluate the free energy of the coplanar state by using Eq. (\ref{eq:GL_final}), it is higher than that of the hedgehog lattice. If the coefficient $A_3$ defined in Eq. (6) were large enough, the coplanar state could be stable, but this is not the case in the present model. Thus, higher-order contributions in the GL free energy, which have not been considered in Eq. (7), or thermal fluctuations may be relevant for the stability of the coplanar phase. One can see from Fig. \ref{fig:snap_comp} (b) that in the coplanar phase, four sublattices represented by different colors are divided into two pairs and spins belonging to each pair have the same polarization axis, suggesting that the spin state is coplanar although the coplanarity is disturbed by the thermal noise.
We note that as we will discuss below, even if this state were realized at a low temperature escaping from the thermal noise, the spins in each pair could not be perfectly collinear.  

The hedgehog-lattice phase is realized at low temperatures in the wide range of the parameter space. Although  the real-space spin configuration at $J_1'/J_1=0.6$ and $T/J_1=0.01$ is presented in Fig. \ref{fig:hedgehog_snapshot}, here, we show the one for the stronger breathing alternation of $J_1'/J_1=0.2$ in Fig. \ref{fig:snap_comp} (c). The snapshot is obtained by cooling the system down to the very low temperature $T/J_1=0.001$ to reduce the thermal noise. One can see from the lower panel of Fig. \ref{fig:snap_comp} (c) that the polarization vector of each sublattice splits into four, although the real-space spin configuration in the upper panel does not differ so much from Fig. \ref{fig:hedgehog_snapshot}. As we will address below, such a splitting is generic to the quadruple-{\bf q} $(\frac{1}{2},\frac{1}{2},\frac{1}{2})$ noncollinear states on the breathing pyrochlore lattice. Actually, when the spin configuration in Fig. \ref{fig:hedgehog_snapshot} is further cooled down, the splitting overcomes the thermal noise and becomes visible. The origin of the splitting can be understood within the MF approximation.

\begin{figure}[t]
\begin{center}
\includegraphics[scale=0.5]{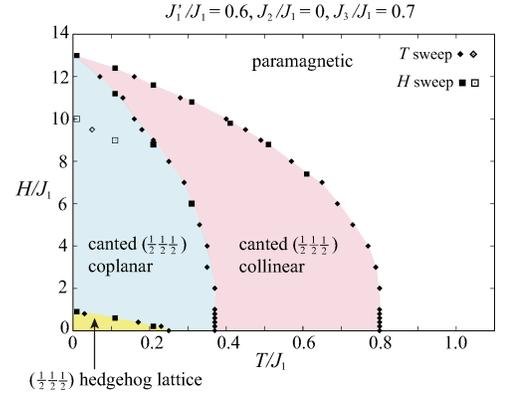}
\caption{The temperature and magnetic-field phase diagram obtained in the MC simulations for $J'_1/J_1=0.6$, $J_2/J_1=0$, and $J_3/J_1=0.7$. \label{fig:HT_J06}}
\end{center}
\end{figure}
Since the Fourier component of spin $\langle {\bf S}^\alpha_{\bf q}\rangle$ is obtained with the combined use of $\mbox{\boldmath $\Phi$}_{{\bf Q}_\mu} = \frac{\overline S}{\sqrt{8}} e^{i \theta_\mu} {\hat P}_\mu$ and Eqs. (\ref{eq:Uq}) and (\ref{eq:S_q}), spins belonging to sublattice $\alpha$, ${\bf S}_j^\alpha$'s, are given by
\begin{eqnarray}
{\bf S}_j^1 &=& \frac{{\overline S}}{\sqrt{2}}\Big[ \hat{P}_4 \cos({\bf Q}_4\cdot{\bf r}_j + \theta_4) +\varepsilon \big\{- \hat{P}_1 \cos({\bf Q}_1\cdot{\bf r}_j + \theta_1) \nonumber\\
&& + \hat{P}_2 \cos({\bf Q}_2\cdot{\bf r}_j + \theta_2) + \hat{P}_3 \cos({\bf Q}_3\cdot{\bf r}_j + \theta_3) \big\} \Big], \nonumber\\
{\bf S}_j^2 &=& \frac{{\overline S}}{\sqrt{2}}\Big[ \hat{P}_1 \cos({\bf Q}_1\cdot{\bf r}_j + \theta_1)  +\varepsilon \big\{ \hat{P}_2 \cos({\bf Q}_2\cdot{\bf r}_j + \theta_2) \nonumber\\
&& + \hat{P}_3 \cos({\bf Q}_3\cdot{\bf r}_j + \theta_3) + \hat{P}_4 \cos({\bf Q}_4\cdot{\bf r}_j + \theta_4) \big\} \Big], \nonumber\\
{\bf S}_j^3 &=& \frac{{\overline S}}{\sqrt{2}}\Big[ \hat{P}_2 \cos({\bf Q}_2\cdot{\bf r}_j + \theta_2) +\varepsilon \big\{- \hat{P}_1 \cos({\bf Q}_1\cdot{\bf r}_j + \theta_1) \nonumber\\
&& + \hat{P}_3 \cos({\bf Q}_3\cdot{\bf r}_j + \theta_3) + \hat{P}_4 \cos({\bf Q}_4\cdot{\bf r}_j + \theta_4) \big\} \Big], \nonumber\\
{\bf S}_j^4 &=& \frac{{\overline S}}{\sqrt{2}}\Big[ \hat{P}_3 \cos({\bf Q}_3\cdot{\bf r}_j + \theta_3) +\varepsilon \big\{- \hat{P}_1 \cos({\bf Q}_1\cdot{\bf r}_j + \theta_1) \nonumber\\
&& + \hat{P}_2 \cos({\bf Q}_2\cdot{\bf r}_j + \theta_2) + \hat{P}_4 \cos({\bf Q}_4\cdot{\bf r}_j + \theta_4) \big\} \Big] .
\end{eqnarray}
As long as $\hat{P}_\mu$'s are not collinear like in the cases of the coplanar and hedgehog-lattice phases, spins on each sublattice cannot be collinear because of the existence of $\varepsilon$. Particularly in the hedgehog-lattice phase, all the four spins on each sublattice within the cubic unit cell orient in the different directions, exhibiting the four spots when the spins are mapped onto the sphere.
Since $\varepsilon$ is nonzero only in the breathing case, such a splitting is inherent to the $(\frac{1}{2},\frac{1}{2},\frac{1}{2})$ noncollinear states on the breathing pyrochlore lattice. We note that $\varepsilon$ is important for the occurrence of the hedgehog-lattice phase as explained in Sec. II, but the resultant splitting itself is not. \\

\section{The temperature and magnetic-field phase diagram in the weaker breathing case of $J'_1/J_1=0.6$}
Figure \ref{fig:HT_J06} shows the temperature and magnetic-field phase diagram for $J'_1/J_1=0.6$, $J_2/J_1=0$, and $J_3/J_1=0.7$. Compared with the strongly breathing case of $J'_1/J_1=0.2$ [see Fig. \ref{fig:HT} (a)], the canted collinear state additionally shows up in the higher temperature region, and the stability region of the hedgehog lattice is suppressed. In the canted coplanar state, a weak transition-like anomaly is also found for $J'_1/J_1=0.6$ as well as $J'_1/J_1=0.2$ (see open symbols in Fig. \ref{fig:HT_J06}).\\

\section{Analytical calculation of the local and total chiralities in the in-field hedgehog-lattice phase}
\begin{figure}[t]
\begin{center}
\includegraphics[scale=0.52]{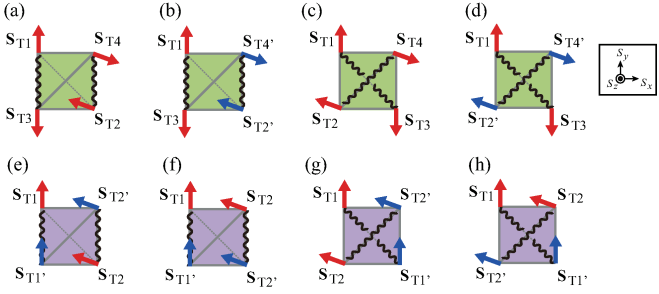}
\caption{Spin configurations of a tetragonal-symmetric tetrahedron projected onto a two-dimensional plane. (a)-(d) AF-type tetrahedra, and (e)-(h) F-type tetrahedra. A red (blue) arrow represents a $S_xS_y$ component of a spin vector pointing upward (downward), and a wavy line denotes a pair bond. \label{fig:schi_isotetra}}
\end{center}
\end{figure}
As shown in Figs. \ref{fig:schi} (a)-(d), when a tetragonal-symmetric tetrahedron is projected onto a two-dimensional plane, two spins in each pair are arranged adjacently or diagonally on the plane. Here, we analytically calculate the chirality on the two-dimensional plane for various spin arrangements of a single tetrahedron shown in Fig. \ref{fig:schi_isotetra}, and apply the result on the single tetrahedron to the in-field hedgehog lattice involving the 8 small tetrahedra.   \\

We first consider the AF-type tetrahedron in which the $S_xS_y$ components in each pair are antiferromagnetically aligned and the $S_z$ ones ferromagnetically. Figures \ref{fig:schi_isotetra} (a)-(d) show the adjacent [(a) and (b)] and diagonal [(c) and (d)] arrangements for the AF-type tetrahedron. Since the scalar chirality is invariant under any rotations with respect to the applied field direction, i.e., $S_z$, we can fix a $S_xS_y$ component of one spin to be, for example, in the $S_y$ direction. Thus, we assume that two spins in one pair are given by
\begin{equation}\left\{\begin{array}{l}
{\bf S}_{\rm T1} = (0, \, S^{+}_{xy}, \, m+\delta S_z) \nonumber\\
{\bf S}_{\rm T3} = (0, \, -S^{+}_{xy}, \, m+\delta S_z) 
\end{array} \right .
\end{equation}
with $S^{\pm}_{xy} = \sqrt{1-(m \pm \delta S_z)^2}$. Then, the rest two spins in the other pair can be expressed as
\begin{equation}\left\{\begin{array}{l}
{\bf S}_{\rm T2} = (S^{+}_{xy} \cos\phi, \, S^{+}_{xy} \sin\phi, \, m+ \delta S_z) \nonumber\\
{\bf S}_{\rm T4} = (-S^{+}_{xy} \cos\phi, \, -S^{+}_{xy} \sin\phi, \, m+ \delta S_z) 
\end{array} \right .
\end{equation}
or 
\begin{equation}\left\{\begin{array}{l}
{\bf S}_{\rm T2'} = (S^{-}_{xy} \cos\phi, \, S^{-}_{xy} \sin\phi, \, m-\delta S_z) \nonumber\\
{\bf S}_{\rm T4'} = (-S^{-}_{xy} \cos\phi, \, -S^{-}_{xy} \sin\phi, \, m- \delta S_z) ,
\end{array} \right .
\end{equation}
where $\phi$ determines the angle between the $S_xS_y$-plane collinear axes of the two pairs. The single-tetrahedron chirality on the two-dimensional plane is given by $\chi^{\rm 2D,tetra}=\sum_{\langle i,j,k \rangle} \chi_{ijk}$, where $\langle i,j,k \rangle$ denotes the summation over all the four triads and $\chi_{ijk}={\bf S}_i \cdot ({\bf S}_j \times {\bf S}_k)$ is defined in a way such that $i$, $j$, and $k$ are ordered in the anticlockwise direction with respect to the surface normal. For the spin configurations shown in Figs. \ref{fig:schi_isotetra} (a)-(d), the single-tetrahedron chirality $\chi^{\rm 2D,tetra}$ can be calculated as
\begin{equation}\label{eq:schi_AFtype}
\chi^{\rm 2D,tetra}=\left\{\begin{array}{l}
0 \qquad  \qquad  \qquad  \qquad  \qquad  \qquad   {\rm [Fig.\, 9 (a)] }  \\
0 \qquad  \qquad  \qquad  \qquad  \qquad  \qquad   {\rm [Fig.\, 9 (b)] }  \\
-8 \, (S^{+}_{xy})^2 \, (m+\delta S_z) \cos\phi \qquad  {\rm [Fig.\, 9 (c)] } \\
-8 \, S^{+}_{xy} \, S^{-}_{xy} \, m  \cos\phi  \qquad \qquad   \quad {\rm [Fig.\, 9 (d)] }  .
\end{array} \right .
\end{equation}

\begin{figure}[t]
\begin{center}
\includegraphics[scale=0.5]{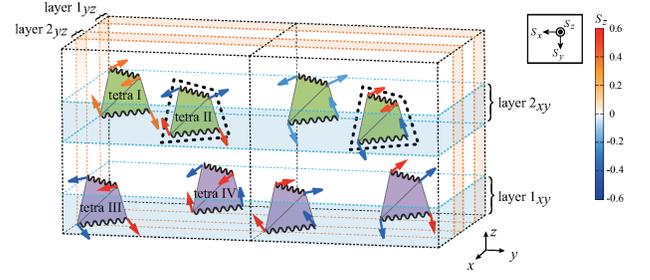}
\caption{Real-space spin configuration associated with Figs. \ref{fig:schi} (g) and (h). The $S_xS_y$ ($S_z$) component of a spin is represented by the arrow (color scale), and other color and symbol notations are the same as those in Figs. \ref{fig:schi} (e) and (f). \label{fig:schi_tetra_real}}
\end{center}
\end{figure}
Next, we consider the F-type tetrahedron in which the $S_xS_y$ components in each pair are ferromagnetically aligned and the $S_z$ ones antiferromagnetically. Figures \ref{fig:schi_isotetra} (e)-(h) show the adjacent [(e) and (f)] and diagonal [(g) and (h)] arrangements for the F-type tetrahedron. In the F-type case, we assume that two spins in one pair are given by
\begin{equation}\left\{\begin{array}{l}
{\bf S}_{\rm T1} = (0, \, S^{+}_{xy}, \, m+\delta S_z) \nonumber\\
{\bf S}_{\rm T1'} = (0, \, S^{-}_{xy}, \, m-\delta S_z) 
\end{array} \right . .
\end{equation}
Then, the rest two spins in the other pair can be expressed as
\begin{equation}\left\{\begin{array}{l}
{\bf S}_{\rm T2} = (S^{+}_{xy} \cos\phi, \, S^{+}_{xy} \sin\phi, \, m+ \delta S_z) \nonumber\\
{\bf S}_{\rm T2'} = (S^{-}_{xy} \cos\phi, \, S^{-}_{xy} \sin\phi, \, m- \delta S_z)  .
\end{array} \right .
\end{equation} 
The single-tetrahedron chirality on the two-dimensional plane $\chi^{\rm 2D,tetra}$ can be calculated as
\begin{equation}\label{eq:schi_Ftype}
\chi^{\rm 2D,tetra}=\left\{\begin{array}{l}
0 \qquad  \qquad  \qquad  \qquad  \qquad  \qquad \qquad    {\rm [Fig.\, 9 (e)] }  \\
-2\Big[ \big\{(S^{-}_{xy})^2-(S^{+}_{xy})^2 \big\} \, m \\
 \quad + (S^{-}_{xy}+S^{+}_{xy})^2 \, \delta S_z \Big] \cos\phi \qquad  {\rm [Fig.\, 9 (f)] } \\
-2\Big[ (S^{-}_{xy}-S^{+}_{xy})^2  \, m \\
\quad + \big\{(S^{-}_{xy})^2-(S^{+}_{xy})^2 \big\} \, \delta S_z \Big] \cos\phi \quad  {\rm [Fig.\, 9 (g)] }  \\
 2\Big[ (S^{-}_{xy}-S^{+}_{xy})^2  \, m \\
\quad + \big\{(S^{-}_{xy})^2-(S^{+}_{xy})^2 \big\} \, \delta S_z \Big] \cos\phi  \quad {\rm [Fig.\, 9 (h)] }  .
\end{array} \right . 
\end{equation}

\begin{figure}[t]
\begin{center}
\includegraphics[scale=0.5]{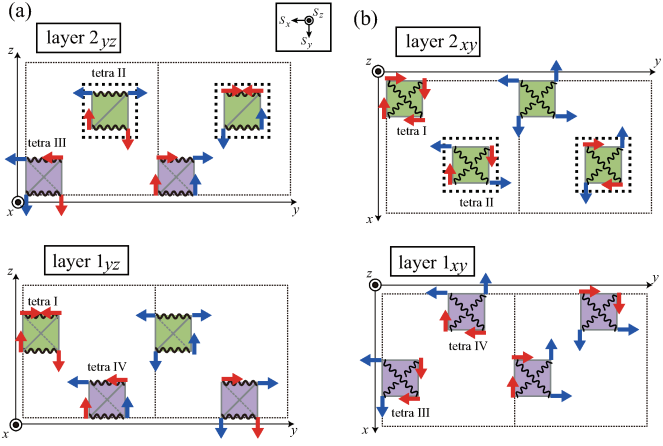}
\caption{Schematically-drawn layer-resolved spin configurations of Fig. \ref{fig:schi_tetra_real}. (a) The simplified spin configuration of Fig. \ref{fig:schi_tetra_real} on the $yz$-plane layers defined in Fig. \ref{fig:schi_tetra_real}, and (b) those on the $xy$-plane layers, where a red (blue) arrow represents the $S_xS_y$ component of a spin vector pointing upward (downward) and other notations are the same as those in Fig. \ref{fig:schi_tetra_real}. In each tetrahedron, the $S_xS_y$ components of the two different pairs are assumed to be orthogonal to each other for simplicity.  \label{fig:schi_tetra}}
\end{center}
\end{figure}
As the chirality for a single tetrahedron $\chi^{\rm 2D,tetra}$ is obtained, we will turn to the total chirality in the hedgehog lattice phase. The spin configuration associated with Figs. \ref{fig:tetragonalspin} and \ref{fig:schi} is shown in Fig. \ref{fig:schi_tetra_real}. Note that the spin structure is tetragonal-symmetric with respect to the $z$-axis. This spin configuration is schematically drawn in the layer-resolved form in Fig. \ref{fig:schi_tetra}, where for simplicity, $\phi$ is assumed to be $0$ or $\pi$ such that the angle between the $S_xS_y$-plane collinear axes of the two pairs be orthogonal to each other. Since the hedgehog lattice consists of the alternating array of the two cubic unit cells having oppositely oriented spins except the uniform magnetization $m$, the two neighboring unit cells on the projected plane (two neighboring unit cells in Fig. \ref{fig:schi_tetra}) involve the contribution given by Eqs. (\ref{eq:schi_AFtype}) and (\ref{eq:schi_Ftype}) and the counter contribution obtained by carrying out the replacement
\begin{equation}\label{eq:replace}
S^{\pm}_{xy} \rightarrow - \, S^{\mp}_{xy}, \quad \delta S_{z} \rightarrow -\, \delta S_{z} 
\end{equation}
in Eqs. (\ref{eq:schi_AFtype}) and (\ref{eq:schi_Ftype}). 

On the plane perpendicular to the non-tetragonal-symmetric direction [layers $1_{yz}$ and $2_{yz}$ in Fig. \ref{fig:schi_tetra} (a)], the paired spins are arranged adjacently, so that $\chi^{\rm 2D,tetra}$ becomes zero for almost all the tetrahedra except tetra III, which corresponds to Fig. \ref{fig:schi_isotetra} (f), and its counter tetrahedron in the neighboring unit cell. The tetra III yields the nonzero $\chi^{\rm 2D,tetra}$ given by Eq. (\ref{eq:schi_Ftype}) for Fig. \ref{fig:schi_isotetra} (f), but it is {\it completely} canceled out by the contribution from the counter tetrahedron because of Eq. (\ref{eq:replace}). Hence, the total chirality on the two-dimensional plane vanishes when the two paired spins are arranged adjacently.

In contrast, on the plane perpendicular to the tetragonal-symmetric direction [layers $1_{xy}$ and $2_{xy}$ in Fig. \ref{fig:schi_tetra} (b)] where the two paired spins are arranged diagonally, the local chirality $\chi^{\rm 2D,tetra}$ becomes nonzero for all the tetrahedra. Since the four tetrahedra in the cubic unit cell [tetra's I, II, III, and IV in Fig. \ref{fig:schi_tetra} (b)] are not equivalent to one another, we will take the $\delta S_z$ components for these tetrahedra, $\delta S_{z,{\rm I}}$, $\delta S_{z,{\rm II}}$, $\delta S_{z,{\rm III}}$, and $\delta S_{z,{\rm IV}}$, as independent variables. After averaging over all the tetrahedra within the two unit cells, we obtain the total chirality per tetrahedron as
\begin{eqnarray}\label{eq:layer1}
&& m \Big[ (S^-_{xy,{\rm III}}-S^+_{xy, {\rm III}})^2 + 4(\delta S_{z,{\rm III}})^2 \nonumber\\
&& \qquad + (S^-_{xy,{\rm IV}}-S^+_{xy,{\rm IV}})^2 + 4(\delta S_{z,{\rm IV}})^2 \Big] 
\end{eqnarray} 
for the layer without the monopoles [layer $1_{xy}$ in Fig. \ref{fig:schi_tetra} (b)] and 
\begin{eqnarray}\label{eq:layer2}
&& 2 m \Big[ (S^-_{xy,{\rm I}})^2+(S^+_{xy,{\rm I}})^2 \nonumber\\
&& \qquad -4(\delta S_{z,{\rm I}})^2 -2 S^-_{xy,{\rm II}}S^+_{xy,{\rm II}}\Big]
\end{eqnarray} 
for the layer with the monopoles [layer $2_{xy}$ in Fig. \ref{fig:schi_tetra} (b)], where $S^\pm_{xy,\lambda}$ ($\lambda=$I, II, III, and IV) is defined by $S^{\pm}_{xy,\lambda} = \sqrt{1-(m \pm \delta S_{z,\lambda})^2}$. 
One can see from Eq. (\ref{eq:layer1}) that the total chirality for the layer {\it without} the monopoles definitely takes a nonzero value. In Eq. (\ref{eq:layer1}), the sign of the total chirality is positive since it is calculated in the specific case of Fig. \ref{fig:schi_tetra} (b), but, in general, it can be positive or negative depending on the spin configuration. Actually, when the spin-space inversion $S_x \rightarrow -S_x$ is performed for all the spins in Fig. \ref{fig:schi_tetra} (b), the state is turned to the other chirality-degenerate state with its energy unchanged because the spin Hamiltonian is invariant under the spin-space inversion, and then, the total chirality changes its sign. On the other hand, on the layer {\it with} the monopoles [see Eq. (\ref{eq:layer2})], the cancellation occurs between the different kinds of tetrahedra, and as a result, the net chirality on this layer becomes small. 
Thus, the overall chirality is dominated by the contribution from the layers without the monopoles, which has actually been confirmed from the numerical data calculated from the MC spin snapshots.
Since as one can see from Eqs. (\ref{eq:layer1}) and (\ref{eq:layer2}), the total chirality becomes non-vanishing only for nonzero $m$, the magnetic field as well as the symmetry reduction to tetragonal is crucial for the emergence of the topological Hall effect in the present system.

\end{document}